# On Quantum Mechanics

**Carlo Rovelli**

*Department of Physics and Astronomy,*
*University of Pittsburgh, Pittsburgh, Pa 15260, USA.*

Pittsburgh December 10, 1993.

To John Wheeler

**Abstract**

We reformulate the problem of the "interpretation of quantum mechanics" as the problem of deriving the quantum-mechanical formalism from a set of simple physical postulates. We suggest that the common unease with taking quantum mechanics as a fundamental description of nature could derive from the use of an incorrect notion - as the unease with the Lorentz transformations before Einstein derived from the notion of observer-independent time.

Following an analysis of the measurement process as described by different observers, we propose a reformulation of quantum mechanics in terms of information theory. We propose three physical postulates, out of which the formalism of the theory can be reconstructed; these are based on the notion of information that systems have about each other. All systems are assumed to be equivalent: no observer-observed distinction, and information is interpreted as correlation.

We then suggest that the incorrect notion that generates the unease with quantum mechanics is the notion of observer-independent state of a system.

# 1. A reformulation of the problem of the "Interpretation of Quantum Mechanics"

In this paper, we discuss a novel point of view about quantum mechanics. This point of view is not antagonistic to current ones, as the "Copenhagen" [Heisenberg 1927, Bohr 1935], consistent histories [Griffiths 1984, Omnes 1988], coherent histories [Gell-Mann and Hartle 1990], many-worlds [Everett 1957, Wheeler 1957, DeWitt 1970], quantum event [Huges 1989] ... interpretations, but rather combines and complements them. The point of view we discuss is based on a critique of a notion generally assumed uncritically: the notion of absolute, or observer-independent, state of a system. We analyze this notion in detail, and observe that the experimental evidence at the basis of quantum mechanics indicates that: different observers may give different descriptions of the same sequence of events. We then argue that the notion of observer-independent state of a system is inadequate to describe the physical world beyond the $\hbar \to 0$ limit, in the same sense in which the notion of observer-independent time is inadequate to describe the physical world beyond the $c \to 0$ limit. We thus consider the possibility of replacing the notion of observer-independent state with a notion that refers to the relation between physical systems.

## 1.1 Unease

The motivation of the present work is the observation that in spite of the 67 years lapsed from the discovery of quantum mechanics, and in spite of the variety of approaches developed with the aim of clarifying its physical content and improving the original formulation, quantum mechanics still maintains a certain level of obscurity. Quantum mechanics is even accused of being unreasonable and unacceptable, even inconsistent, by world-class physicists (For example, [Newman 1993]). Our point of view in this regard is that quantum mechanics synthesizes most of what we have learned so far about the physical world: The issue is not to replace it, but to understand *what* it actually says about the world, or, equivalently, what precisely we have learned from experimental microphysics.

Still, it *is* difficult to overcome the sense of unease that quantum mechanics communicates. The source of this unease is different in different interpretations. It is expressed, for instance, in the objections the supporters of each interpretation raise against other interpretations. For instance: Is there something "physical" happening when the wave function "collapses"? Is it really possible that observer and measurement, including wave function reduction, could not be described in Schrödinger evolution-terms? How can a classical world emerge from a quantum reality? If somebody would prepare me in a quantum superposition of two ordinary states, how would I feel? If the Planck constant were 25 orders of magnitude larger, how would the world look like? Who chooses the family of consistent histories that describe a set of events, and can this "chooser" be described quantum mechanically? And so on. Some of these questions are probably naive, or ill posed. But the fact that they keep being asked is, we believe, a sign that the problem of the interpretation of quantum mechanics has not been fully disentangled yet. In this paper, we address this sense of unease.

## 1.2 A strategy and a historical model

Our strategy is characterized by two ideas:
1. We take a different route to the problem of giving an interpretation to quantum mechanics than most of the current interpretations: The effort in this paper is not to *append* a reasonable interpretation to the quantum mechanics *formalism*, but rather to *derive* the formalism from a set of physical, experimentally motivated, assertions ("postulates", or "principles") about the world. We suspect that such a derivation is what may unravel the puzzle of the interpretation of quantum mechanics.
2. We suggest that the sense of unease with quantum mechanics derives from the use of a concept inappropriate to describe the physical world. We argue that this concept is the concept of observer-independent state of a system.
The reasons for exploring such a strategy are illuminated by considering an obvious historical precedent: special relativity.

Special relativity is a well understood physical theory, accredited to Einstein's 1905 celebrated paper. However, the formal content of special relativity is entirely coded in the Lorentz transformations, which were written by Lorentz, not by Einstein, and several years before 1905. What was Einstein's contribution? It was to understand the physical meaning of the Lorentz transformations. We could say, in a provocative manner, that Einstein's contribution to special relativity was the *interpretation* of the theory, not its *formalism* : the formalism already existed. Einstein was so persuasive with his interpretation of the Lorentz equations because he did not *append* an interpretation to them: rather, he *re-derived* them, starting from two "postulates" with clear physical meaning (equivalence of inertial observers - universality of the speed of light) taken as facts of experience.   This procedure unraveled the physical content and provided a solid *interpretation* of the Lorentz transformations.   We suggest that, in order to grasp the full physical meaning of quantum mechanics, a similar result should be achieved: Find a small number of simple statements with clear physical meaning, from which the formalism of quantum mechanics could be *derived*.   To our knowledge, such a derivation has not been achieved yet.   In this paper, we do not reach this result in a fully satisfactory manner, but we discuss a possible reconstruction scheme.

Furthermore, Lorentz transformations were perceived as rather unreasonable and inacceptable, even inconsistent,  before 1905. Lorentz's physical interpretation (physical contraction of bodies moving with respect to the ether, due to complex and unknown electromagnetic interaction between the atoms of the bodies and the ether) was quite unattractive, and remarkably similar to certain interpretations of the wave function collapse presently investigated.   In deriving the Lorentz transformations, Einstein could point out the reason for the unease: the implicit use of a concept (observer-independent time) inappropriate to describe reality, or, equivalently, a common assumption about reality (simultaneity is observer-independent) which was physically incorrect.   The unease with the Lorentz transformations derived from a conceptual scheme in which an *incorrect notion*, absolute simultaneity, was assumed, yielding in any sorts of paradoxical situations.

In this paper, we consider the hypothesis that all "paradoxical" situations associated with quantum mechanics (as the famous and unfortunate half-dead Schrödinger cat [Schrödinger 1935]) follow from some analogous *incorrect notion*  that we use in thinking about quantum mechanics. (Not in using quantum mechanics, since we seem to have learned to use it in a remarkably effective way.)   The main aim of this paper, therefore, is to hunt for this *incorrect notion*, with the hope that by digging it out, and exposing it clearly to everybody's contempt, we could free ourselves from our unease with our best theory of motion, and fully understand what the theory is saying about the world. We will argue that this notion is the notion of observer-independent state of the system.

The program we have outlined is thus to do for the formalism of quantum mechanics what Einstein did for the Lorentz transformations: i. Find a set of simple assertions about the world, with clear physical meaning, that we know are experimentally true (postulates); ii. Analyze these postulates, and show that from their conjunction it follows that certain common assumptions about the world are incorrect; iii. Derive the full formalism of quantum mechanics from these postulates. We expect that if this program could be completed, we would have "understood" quantum mechanics.

## 1.3 Structure of the paper

We analyze the measurement process *as described by two distinct observers*, in section 2. This analysis leads us to the main idea, the observer dependence of the state, and to recognize a few key concepts in terms of which, we suggest, the quantum mechanical description of reality "makes sense". Prominent among these is the concept of information [Shannon 1949, Wheeler 1988, 1989, 1992]. In section 3 we switch from an inductive to a (mildly) deductive mode: we put forward a set of notions, and a set of simple physical statements, from which we reconstruct the formalism of quantum mechanics. We denote these statements as "postulates", at the risk of misunderstanding: we do not claim any mathematical nor philosophical rigor or completeness in our derivation - supplementary assumptions are made along the way. We are not interested here in a formalization of the subject, but only in grasping its "physics". Finally, in section 4 we discuss the picture of the physical world that has emerged. In particular, we compare the approach we have developed with several currently popular interpretations of quantum mechanics, and we argue that the differences between them disappear if one takes into account the analysis presented here.

## 2. Quantum mechanics is a theory about information

In this section, we present a preliminary analysis of the process of measurement, and we introduce the main new ideas. Throughout this section, standard quantum mechanics and standard interpretation (say: formalism and interpretation in [Dirac 1930] or [Messiah 1958]) are assumed.

### 2.1 The third man problem

Consider an observer O that makes a measurement on a system S.  We may think of O (for the moment) as a classical macroscopic measuring apparatus, including or not including a human being - this being irrelevant for what follows.  Assume that the quantity being measured, say Q, may take two values, say A and B (when Q has value A with certainly, we also say: "S has the property A"); and let the states of the system S be described by vectors (rays) in a two (complex) dimensional Hilbert space $H_S$.  Let the two eigenstates of the operator corresponding to the measurement of Q be |A> and |B>.  If, at a time $t=t_1$ prior to the measurement, S is in one of the two eigenstates of Q, say |A>, then, when the measurement is performed, the measuring apparatus detects the value A, and the state of the system S is not affected.  On the other hand, if S is in a generic normalized state  $|\psi> = \alpha$ |A> + $\beta$ |B>, where $\alpha$ and $\beta$ are complex numbers and $|\alpha|^2 + |\beta|^2 = 1$, then O can measure either one of the two values A and B - the respective probabilities being $|\alpha|^2$ and $|\beta|^2$.

Let us assume that in a *given specific measurement*  the outcome of the measurement is A.  From now on, we concentrate on describing *this*  specific experiment, which we denote as **E**.  The system S is affected by the measurement, and at a time $t=t_2$  after the measurement, the state of the system is |A>.  In the physical sequence of events **E**, the states of the system at $t_1$ and $t_2$ are thus

$$t_1 \qquad\qquad \rightarrow \qquad\qquad t_2$$

$$\alpha\,|A>+\beta\,|B> \qquad \rightarrow \qquad\qquad |A> . \qquad\qquad\qquad (1)$$

This is the standard account of a measurement, according to quantum mechanics.

Let us now consider this same sequence of events **E**, as described by a *second*  observer, which we refer to as O'.  O' is an observer different from O.  We refer to O as "he" and to O' as "she".  O' describes a system formed by S and O.  Therefore she views both S and O as sub-systems of the larger S-O system she is considering. Again, we assume O' uses conventional quantum mechanics. We assume that O' does not perform any measurement on the S-O system during the $t_1$-$t_2$ interval, but that she knows the initial states of both S and O, and is able to give a quantum mechanical description of the set of events **E**.    She describes the system S by means of the Hilbert space $H_S$ considered above, and O by means of a Hilbert space $H_O$.    The S-O system is then described by the tensor product $H_{SO} = H_S \otimes H_O$.   Let us denote the vector in $H_O$ that describes the state of the observer O at $t=t_1$ (prior to the measurement) as |init>.    The physical process during which O measures the quantity Q of the system S implies a physical interaction between O and S. In the process of this interaction, the state of O changes.   If the initial state of S is |A> (and the initial state of O is |init>), then |init> evolves to a state that we denote as |OA>.  We may think of |OA> as a state in which the position of the hand of the measuring apparatus points towards the mark "A".  In the course of the interaction S remains in the state |A>. Thus, the state  |A>$\otimes$|initial>  in $H_{SO}$  evolves to the state  |A>$\otimes$|OA>:

$$t_1 \qquad\qquad\qquad \rightarrow \qquad t_2$$
$$|A>\otimes\ |init> \qquad \rightarrow \qquad |A>\otimes|OA>. \qquad\qquad (2)$$

Analogously, if the initial state of the S-O system is  |B>$\otimes$|init>, it will evolve to a state  |B>$\otimes$|OB>, where |OB> is a state of O that can be thought of as describing the position of the hand of the measuring apparatus pointing towards the mark "B"

$$t_1 \qquad\qquad\qquad \rightarrow \qquad t_2$$
$$|B>\otimes\ |init> \qquad \rightarrow \qquad |B>\otimes|OB>. \qquad\qquad (3)$$

It is not difficult to construct model Hamiltonians that produce evolutions of this kind, and can be taken as models for the physical interactions that produce such a measurement.

Let us now consider the actual case of the experiment **E**, in which the initial state of S is $|\psi> = \alpha\ |A>+\beta\ |B>$. The initial full state of the S-O system is then

$$|\psi>\otimes|init> \ = (\alpha\ |A>+\beta\ |B>)\otimes\ |init>\ .$$

Because of the linearity of quantum mechanics, we have from (2) and (3):

$$t_1 \qquad\qquad\qquad\qquad \rightarrow \qquad t_2$$
$$(\alpha\ |A>+\beta\ |B>)\otimes|init> \quad \rightarrow \quad \alpha\ |A>\otimes|OA> + \beta\ |B>\otimes|OB>. \quad (4)$$

Thus, at t=$t_2$ the system S-O is in the state $\alpha$ |A>⊗|OA>+$\beta$ |B>⊗|OB>. This is the conventional description of a measurement as a physical process, originally due to Jauch [Jauch 1968].

We have described an actual physical process **E** taking place in a real laboratory. Standard quantum mechanics requires us to distinguish system from observer, but it allows us some freedom in drawing the line that distinguishes the two [von Neumann 1932]. The peculiarity of the above analysis is just the fact that we exploited this freedom in order to describe the same sequence of physical events in terms of two different descriptions. In the first description, the line that distinguishes system from observer is set between S and O. In the second , between S-O and O'. We recall that we assume that O' is not making a measurement on the S-O system; there is no physical interaction between S-O and O' during the $t_1$-$t_2$ interval. (O' may perhaps make a measurement at a later time $t_3$: then, as shown by von Neumann, the probabilities and the final quantum states that one computes assuming that the collapse happened between $t_1$ and $t_2$ or between $t_2$ and $t_3$ are in agreement.)

Thus, we have two descriptions of the physical sequence of events that we have denoted **E**: The description (1) given by the observer O and the description (4) given by the observer O'. The point we would like to stress here is that these are two distinct *correct* descriptions of the same sequence of events **E**. In the O description, the system S is in the state |A> and the hand of the measuring apparatus indicates "A". While according to the O'-description, S is not in the state |A> and the hand of the measuring apparatus does not indicate "A".

One can take the view that outcomes of measurements are the only physical content of the theory [Heisenberg 1927, van Fraassen 1991]. Then the account (4) has nothing to say about reality at time $t_2$, and O' does not know anything about what happens between $t_1$ and some later time $t_3$ in which she (O') performs a measurement; anything in between is like the "non-existing" trajectory of the electron, of which we are not allowed to say anything. However, even within this perspective, O and O' give very different accounts of S at t=$t_2$: O claims that, at t=$t_2$, S has the property A; while O' claims that at t=$t_2$ the quantity Q does not have any determined value. Thus: the two accounts differ both as far as states are concerned and as far as measurements' outcomes are concerned. The conclusion we derive

from this discussion, and that we are going to explore in the rest of this work, is then:

   *Main observation:* *In quantum mechanics different observers may give different accounts of the same sequence of events.*

## 2.2 Relation between the two descriptions

Does the observer O' "know" that O has made a measurement on S? The answer is yes, for a number of reasons. First, the observer O' has a full account of the events in **E**, so it should be possible to interpret her description (4) as expressing the fact that O has measured S. Indeed this is possible: The key observation is that the final state at $t_2$ is a correlated state. This state is given as the quantum superposition of two states: in the first, ($|A>\otimes|OA>$), S is in the $|A>$ state and the hand of the observer is on the "A" mark. In the second, ( $|B>\otimes|OB>$), S is in the $|B>$ state and the hand of the observer is on the "B" mark. In both cases, the hand of O is on the mark that correctly represents the state of the system. More technically, O' could measure an observable M that checks whether the hand of O indicates the correct state of S. If he did so, then the outcome of this measurement would be "yes" with certainty. The operator M is given by

   $M |A>\otimes|OA> = |A>\otimes|OA>,$          $M |A>\otimes|OB> = 0 ,$

   $M |B>\otimes|OB> = |B>\otimes|OB> ,$          $M |B>\otimes|OA> = 0 ,$          (5)

where the eigenvalue 1 means "yes, the hand of O indicates the correct state of S" and the eigenvalue 0 means "no, the hand of O does not indicates the correct state of S". Thus, it is perfectly meaningful to say that, according to the O' description, O "knows" the quantity Q of S, or that he "has measured" the quantity Q of S.

On the other side, the observer O' does not know "what is the value of the observable Q". (She could measure it, but the outcome of the measurement is not determined with certitude at $t=t_2$.) In particular she does not know "what precisely O knows about Q". (Again, she could measure whether O is in $|OA>$, or in $|OB>$.) Thus, we must distinguish the statement "O' knows that O knows about Q" from the statement "O' knows what O knows about Q". Of course this is a consistent distinction, common in everyday life (I know *that* you know the amount of your salary, but I do not know *what* you know about the amount of your salary).  This distinction will soon become important.

The feature of the final state $\alpha\, |A\rangle\otimes|OA\rangle+\beta|B\rangle\otimes|OB\rangle$ that allows us (or O') to interpret it as containing the information that O has measured S is the fact that the state is a correlated state. The operator M, indeed, essentially checks the correlation between $|A\rangle$ and $|OA\rangle$ and between $|B\rangle$ and $|OB\rangle$.  Thus, from the point of view of the O' description:

> *The fact that* O *knows about* S *(or has measured* Q*) is expressed by the existence of a correlation between the state of* S *and the state of* O.

## 2.3   Main discussion

Let us return to the "Main observation" that we stated at the end of section 2.1, namely the observation that the description of a system is observer dependent.

A possibility of avoiding such a conclusion is to deny the possibility that O, an observer, could be described as a quantum system.  This is Bohr's point of view [Bohr 1949]. If we accept it, then we have to separate reality into two kinds of systems: quantum mechanical systems on one side, and classical observers (or a unique observer) on the other side. Bohr claims explicitly that we have to renounce giving a full quantum mechanical description of the classical world [Bohr 1949, Landau and Lifschit 1977]. Wigner pushes this view to the extreme consequences and distinguishes material systems (observed) from consciousness (observer) [Wigner 1961].  Here, on the contrary, we wish to assume

> Hypothesis 1: All systems are equivalent: Nothing a priori distinguishes observer systems from quantum systems.  If the observer O can give a description of the system S, then it is also legitimate for an observer O' to give a quantum description of the system formed by the observer O.

In appendix A we discuss other possible objections to the "Main observation"; we also argue that this observation holds also within the "coherent histories" interpretations of the theory.

We may now proceed to the main point of this work:

If, in quantum mechanics, different observers may give different accounts of the same sequence of events, then each quantum mechanical description has to be understood as relative to a particular observer. Thus, a quantum mechanical description of a certain system (state and/or value of measured quantities) should not be taken as an "absolute" (observer independent) description of reality, but rather as a formalization, or codification, of the knowledge, or, more precisely, of the information, that an observer has about a system. Quantum mechanics can therefore be viewed as a theory about the relative information that systems have about each other. These families of information are all related to each other: As illustrated in section 2.2, the fact that a system O has information about a system S is represented as a correlation between the state of S and the state of O, when viewed as information available to a third system O'.

If quantum mechanics describes only relative information, one could consider the possibility that there is a deeper "underlying" theory that describes what happens "in reality". This is the thesis of the incompleteness of quantum mechanics (First suggested in [Born 1926] !). Examples of such hypothetical "underlying" theories are hidden variables theories [Bohm 1951, Belifante 1973]. Alternatively, the "wave-function-collapse-producing" systems can be "special" because of some non-yet-understood physics, which becomes relevant due to large number of degrees of freedom [Ghirardi Rimini and Weber 1986, Bell 1987], complexity [Hughes 1989], or quantum gravity [Penrose 1989]. As is well known, there are no indications on *physical* grounds that quantum mechanics is incomplete. Indeed, the *practice* of quantum mechanics supports the view that quantum mechanics represents the best we can say about the world at the present state of experimentation, and suggests that the structure of the world grasped by quantum mechanics is deeper, and not more approximate, than the scheme of description of the world provided by classical mechanics. But one could consider motivations on *metaphysical* grounds, in support to the incompleteness of quantum mechanics. One could argue: "Since reality has to be real and universal, and the same for everybody, then a theory in which the description of reality is observer-dependent is certainly an incomplete theory". If such a theory were complete, our concept of reality would be disturbed.

However, the way we have reformulated the problem of the interpretation of quantum mechanics in sec.1 should make us suspicious and attentive to such kinds of arguments. Indeed, what we are looking for is *precisely* some "wrong assumption" that we suspect to have, and that could be at the origin of the difficulty in understanding quantum mechanics. Thus, we discard here the thesis of the incompleteness of quantum mechanics and assume

Hypothesis 2: (Completeness) Quantum mechanics provides a complete and self-consistent scheme of description of the physical world, appropriate to our present level of experimental observations.

The conjunction of this hypothesis 2 with the "Main observation" of section 2.1 and the discussion above leads us to the following idea:

*Quantum mechanics is a theory about the relative information that sub-systems have about each other, and this is a complete description of the world.*

The main thesis of this paper is that this conclusion is not self-contradictory. If this conclusion is valid, then we have found the "incorrect notion" at the source of our unease with quantum theory: The notion of "the 'true' or observer-independent description of the world". If a complete description of the world is exhausted by the relevant information that systems have about each other, then there is no observer-independent description of the world. Namely, there is neither an absolute "state of the system", nor an absolute "property that the system has at a certain time". The accounts (1) and (4) of the sequence of events **E** are both correct, even if distinct: any time we talk about a "state" or "property" of a system, we have to refer these notions to a specific observing system.

Thus, we suggest that quantum mechanics indicates that the notion of a "universal" description of the the state of the world, shared by all observers, is a physically misleading concept. A *description* of the state of a system S exists only if somebody or something (namely another system considered as an *observer*) is actually "describing" S, or, more precisely, has information about S. The state of a system is always a state of that system with respect to a certain observer [Crane 1991]. A general physical theory is a theory about the information that observers have about systems. We explore and elaborate this possibility in this paper.

## 2.4    Information

The relevance of the notion of information for understanding the deep nature of quantum physics has been advocated by John Wheeler [Wheeler 1988, 1989, 1992].    The notion of information we employ here should not be confused with other notions of information employed in other contexts. We elucidate this notion from two complementary perspectives.

First, if we collect a certain amount of data about the world, and organize them within some theoretical structure, we have collected a certain amount of information.    Let us distinguish the notion of "theory about the world" on one side from the notion of "theory about our knowledge of the world", or "theory about the information we have about the world" on the other side. This distinction is irrelevant if we assume that a "universal" description of the world exists and is shared by all observers.   (A theory with a notion of "time for an observer" as opposite to just "time" is irrelevant if simultaneity is absolute.)   But if different observers give different descriptions of the same system, then the distinction becomes relevant.

This notion of information, however, is too vague for a physical theory. We need a simpler and more physically precise notion that characterizes the information that a system has: a notion that does not require us to distinguish between human and no-human observers, systems that understand meaning or that don't, very-complicated or simple systems, or so on.   As it is well known, this problem was brilliantly solved by Claude Shannon.   The technical notion of amount of information that we need is the one from information theory [Shannon 1949]. In the technical sense of information theory, information is simply contained in the fact that a system is in a certain configuration, which is (has been) correlated to some information source;   the amount of information is the number of the elements of a set of alternatives out of which the configuration is chosen.   The relation between this notion of information and more elaborate uses of the notion of information is simply given by the fact that existence of this information-theoretical information is a necessary condition for any "elaborate information" to exist.   Thus, for a physical theory it is sufficient to deal with this information-theoretical notion of information.

More precisely, the analysis in sec. 2.2 has indicated the following: The fact that O has information about S simply means (when seen from O') that the states of O and S are correlated, or that the state of the S-O system (as described by O') is an eigenstate of the M operator. This notion of information is a very weak notion, and does not require us to consider information storage, thermodynamics, complex systems, meaning, or anything of the sort. In particular, according to this use of the notion of information: i. information can be lost dynamically (correlated state may become uncorrelated); ii. we do not distinguish between "obtained" correlation and "accidental" correlation (if the pointer of the apparatus indicates the correct value of the spin, we say that the pointer has information about the spin, whether or not this is the outcome of a "well thought" interaction); ii. any physical system can contain information about another physical system. For instance if we have two spin-1/2 particles that have the same value of the spin in the same direction, we say that one has information about the other one.

This last point is crucial because if there is any hope of understanding how a system may behave as an observing system, without renouncing the postulate that all systems are equivalent, then the same kind of processes ("collapse") that may happen between an electron and a CERN machine, may also happen between an electron and another electron. Observers are not "physically special systems".

In the next section, we make use of the concepts that emerged from this discussion, we assume a certain number of postulates expressed in terms if these concepts, and we derive quantum mechanics from these postulates.

## 3.    Reconstruction  of  Quantum  Mechanics

## 3.1 Basic  concepts

The physical theory is concerned with relations between physical systems.  In particular, it is concerned with the description that observers give about observed systems. Following our hypothesis 1, we reject any fundamental or metaphysical distinctions as:  system / observer,  quantum system / classical system,  physical system / consciousness.  We assume the existence of an ensemble of systems, each of which can be equivalently considered as an *observing system* or as an *observed system*.  A system (*observing system*) may have *information* about another system (*observed system*).  Information is exchanged via physical interactions. The actual process through which information is collected and stored is not of particular interest here, but can be physically described in any specific instance.

A key idea from information theory that plays a role here is the fact that information is a discrete quantity: there is a minimum amount of information exchangeable (a single bit, or the information that distinguishes between just two alternatives.)   The process of acquisition of information (a measurement) can be described as a "question" that a system (observing system) asks  another  system (observed system).   Since information is discrete, any process of acquisition of information can be decomposed into acquisitions of elementary bits of information.  We refer to an elementary question that collects a single bit of information as a "yes/no question", and we denote these questions as $Q_1$, $Q_2$, ... .

We assume that any system S, viewed as an observed system, is characterized by a family of yes/no questions that can be meaningfully asked to it. We denote the set of these questions as $\amalg$(S) = {$Q_i$, i∈ I}, where the index i belongs to a set I (characteristic of S).  The general kinematical features of S are representable as relations between the questions $Q_i$ in $\amalg$(S), that is, as structures over $\amalg$(S).  For instance, meaningful questions that can be asked to an electron are whether the particle is in a certain region of space, whether its spin along a certain direction is positive, and so on.

By asking a sequence of questions ($Q_1$, $Q_2$ , $Q_3$ ... ) to S, an observer system O may compile a string

$$(e_1, e_2, e_3, \ldots \ldots) \tag{6}$$

where each $e_i$ is either 0 or 1 ("no" or "yes") and represents the "answer" of the system to the question $Q_i$. (More precisely, the information that O has about S can be represented as a string.)  It is of course a basic fact about nature that knowledge of a portion $(e_1, \ldots, e_n)$ of this string provides indications about the subsequent outcomes $(e_n, e_{n+1}, \ldots)$. It is in this sense that a string (6) contains the information that O has about S.

We distinguish between information contained in an arbitrary string (6) and *relevant* information. Repeating the same question (experiment) and obtaining always the same outcome does not increase the information on S. The *relevant information* (sometimes simply *information*) that O has about S is defined as the non-trivial content of the (potentially infinite) string (6), that is the part of (6) relevant for predicting future answers of possible future questions. The relevant information is a subset of the string (6), obtained discarding the $e_i$'s that do not affect the outcomes of future questions.

Let us make a few observations on the relation between the notions introduced and traditional notions used in quantum mechanics.
- The notion of question is of course a version of the notion of measurement.  The idea that any measurement in quantum mechanics can be reduced to yes/no measurements is well known. A yes/no measurement is represented by a projection operator onto a linear subset of the Hilbert space, or, equivalently, by the linear subset of the Hilbert space itself.  In the present framework this idea is not derived from the quantum mechanical formalism, but is justified in information-theoretical  terms.
- The notions of observing system and observed system reflect the traditional notions of observer and system; however, we drastically prevent ourselves from using any sort of distinction between classical and quantum systems of the sort advocated by Bohr and by several other interpretations of quantum mechanics.
- The set of questions that can be asked about a given system S, namely Ш(S), reflects the notion of the set of the observables.  We recall that in algebraic approaches a system is characterized by the (algebraic) structure of the family of  its observables.
- There is a notion that we have not mentioned here: the *state* of the system.  The absence of this notion is the prime feature of the interpretation that we are introducing.  In place of the notion of state, which refers solely to the system, we have introduced the notion of the information that a system has about another system. We stress the fact that we do not view the notion of information as "metaphysical", but as a concrete notion: we imagine a piece of paper on which outcomes of measurements on S are written, or hands of measuring apparata, or memory of scientists, or a two valued variable which is "up" or "down" after an interaction. We discuss below the precise way in which the existence of this information, and the physical  process through which information is gained, can be self-consistently described in physical terms within the formalism.  As we will see later, this means that a *third*  observer can give a complete account (may have complete information) about the existence of information and the process through which a system acquires  information.

For simplicity, in the following we focus only on the kind of systems that can be described in conventional quantum mechanics by means of finite dimensional Hilbert spaces. (We may imagine that all physical systems can be fully approximated by a discretization.) This choice drastically simplifies the mathematical treatment of the theory (avoiding continuum spectrum and other infinitary issues).  We leave to future work the extension of the present construction to continuum systems.

## 3.2  Postulates

Now, we may introduce the first postulate.

  Postulate 1 (Limited information).   There is a maximum amount
  of *relevant information*   that can be extracted from a system.

The physical meaning of Postulate 1 is that any system is essentially finite, and it is possible to exhaust, or "give a complete description of the system", in a finite time. In other words, any future prediction that can be inferred about the system out of an infinite string (6), can also be inferred from a finite subset

$$s = [e_1, \dots , e_N] \tag{7}$$

of (6), where N is a number that characterizes the system S. The finite string (7) represents the knowledge that O has about S.  We may say that any system S has a maximal "information capacity" N, where N, being an amount of information, is expressed in bits.  This means that N bits of information will exhaust everything we can say about the system S. Thus, each system is characterized by a number N. In terms of traditional notions, N is given by

$$N = \log_2 k, \qquad\qquad (8)$$

where k is the dimension of the Hilbert space of the system S. In a system described by a k dimensional Hilbert space, if I measure a complete set of commuting observables, I can distinguish one possible outcome out of k alternatives (the number of orthogonal basis vectors): in information-theory language, this means that I gain information $N = \log_2 k$ on the system. (We assume here for simplicity that N is an integer; a generalization to non-integer cases should not be difficult.) Postulate 1 is confirmed by our experience about the world (within the assumption above, that we restrict to finite dimensional Hilbert space systems. Generalization to infinite systems should not be difficult). Let us analyze some of its consequences.

In general, the number of questions that one may include in Ш(S) can be much larger than N. Some of these questions may not be independent. In particular, one may find (experimentally) that
i. A question $Q_2$ may depend on a question $Q_1$; namely, it may be that a "yes" answer to $Q_1$ always implies a "yes" answer to $Q_2$. We denote this relation between $Q_1$ and $Q_2$ as

$$Q_1 \Rightarrow Q_2. \qquad\qquad (9)$$

The relation $\Rightarrow$ is reflexive ($Q \Rightarrow Q$), transitive (if $Q_1 \Rightarrow Q_2$ and $Q_2 \Rightarrow Q_3$, then $Q_1 \Rightarrow Q_3$) and, after obvious identifications, antisymmetric (If $Q_1 \Rightarrow Q_2$ and $Q_2 \Rightarrow Q_1$, then $Q_1 = Q_2$). Thus Ш(S) has the structure of a partially ordered set. This is part of the structure of Ш(S) that we mentioned above. This structure is what characterizes S.

ii. A question $Q_3$ may be fully determined by a question $Q_1$ and a question $Q_2$, in the sense of being the minimal question for which $Q_1 \Rightarrow Q_3$ and $Q_2 \Rightarrow Q_3$. (By minimal we mean that if $Q_1 \Rightarrow Q_4$ and $Q_2 \Rightarrow Q_4$, then $Q_3 \Rightarrow Q_4$.) We denote this possibility as

$$Q_3 = Q_1 \vee Q_2. \qquad\qquad (10)$$

iii. Conversely, a question $Q_3$ may have the property that $Q_3 \Rightarrow Q_1$ and $Q_3 \Rightarrow Q_2$, and being maximal, in the sense that if $Q_4 \Rightarrow Q_1$ and $Q_4 \Rightarrow Q_2$, then $Q_4 \Rightarrow Q_3$. We denote this case as

$$Q_3 = Q_1 \wedge Q_2. \tag{11}$$

iv. We may always include in our set of questions a question $Q_o$ whose answer is always "no" and a question $Q_\infty$ whose answer is always "yes". For any $Q_i$, we then have trivially

$$Q_i \Rightarrow Q_\infty, \quad Q_o \Rightarrow Q_i. \tag{12}$$

v. For any question Q, we can introduce a question not-Q, whose answer is "yes" if the answer to Q is "no" and vice versa.

vi. We can introduce a notion of orthogonality as follows. If $Q_1 \Rightarrow$ not-$Q_2$, then we say that $Q_1$ and $Q_2$ are orthogonal, and we indicate this as $Q_1 \perp Q_2$.

Equipped with the structures described in i.-vi., and under the (non-trivial) additional assumption that $\vee$ and $\wedge$ are defined for every pair of questions, ⧢(S) is an orthomodular lattice [Beltrametti and Cassinelli 1981, Hughes 1989]. Other, more subtle ways, in which the outcome of a question may affect the outcome of another question are considered below.

If there is a maximal amount of information that can be extracted from the system, we may assume that out of ⧢(S) one can select an ensemble of N questions $Q_i$, which we denote as ꙟ = {$Q_i$, i=1, N}, that are fully independent one from each other. We do not assume there is anything canonical in this choice, so that there may be many distinct families ꙟ, ꙅ, ꙋ, ... of N independent questions in ⧢(S). If a system O asks the N questions in the family ꙟ to a system S, then the answers obtained can be represented as a string that we denote as

$$s_\text{ꙟ} = [e_1, \ldots , e_N]_\text{ꙟ} \tag{13}$$

The string $s_\text{ꙟ}$ represents the information that O has about S, as a result of the interaction that allowed it to ask the questions in ꙟ. The string $s_\text{ꙟ}$ can take $2^N = k$ values; we denote these values as $s_\text{ꙟ}{}^{(1)}$, $s_\text{ꙟ}{}^{(2)}$, ... ,$s_\text{ꙟ}{}^{(k)}$. So that

$$s_{\mathfrak{q}}^{(1)} = [0, 0, \ldots , 0]_{\mathfrak{q}},$$
$$s_{\mathfrak{q}}^{(2)} = [0, 0, \ldots , 1]_{\mathfrak{q}},$$
$$\ldots$$
$$s_{\mathfrak{q}}^{(k)} = [1, 1, \ldots , 1]_{\mathfrak{q}}. \qquad (14)$$

Since the $2^N$ possible outcomes $s_{\mathfrak{q}}^{(1)}$, $s_{\mathfrak{q}}^{(2)}$, ... $s_{\mathfrak{q}}^{(k)}$ of the N yes/no questions are (by construction) mutually exclusive, we can define $2^N$ new questions $Q_{\mathfrak{q}}^{(1)}$... $Q_{\mathfrak{q}}^{(k)}$ such that the "yes" answer to $Q_{\mathfrak{q}}^{(i)}$ corresponds to the string of answers $s_{\mathfrak{q}}^{(i)}$.   For instance:

$$Q_{\mathfrak{q}}^{(1)} = \text{not } Q_1 \ \Lambda \ \text{not } Q_2 \ \Lambda \ \ldots \ \Lambda \ \text{not } Q_N$$
$$Q_{\mathfrak{q}}^{(2)} = \text{not } Q_1 \ \Lambda \ \text{not } Q_2 \ \Lambda \ \ldots \ \Lambda \ Q_N$$
$$\ldots$$
$$Q_{\mathfrak{q}}^{(k)} = Q_1 \ \Lambda \ Q_2 \ \Lambda \ \ldots \ \Lambda \ Q_N. \qquad (15)$$

We refer to questions of this kind as "complete questions". By taking all possible unions of sets of complete questions $Q_{\mathfrak{q}}^{(i)}$ (of the same family $\mathfrak{q}$), we can construct a Boolean algebra that has $Q_{\mathfrak{q}}^{(i)}$ as atoms.

Alternatively, the observer O could use a *different* family of N independent yes-no questions, in order to gather information about S.  Let us denote this other set as $\mathfrak{o}$.  Then, he will still have a maximal amount of relevant information about S formed by an N-bits string

$$s_{\mathfrak{o}} \ = \ [e_1, \ldots , e_N]_{\mathfrak{o}} \qquad (16)$$

Thus, O can give different kinds of descriptions of S, by asking different questions.  Correspondingly, we denote as $s_{\mathfrak{o}}^{(1)}$... $s_{\mathfrak{o}}^{(N)}$ the $2^N$ values that $s_{\mathfrak{o}}$ can take, and we consider the corresponding complete questions $Q_{\mathfrak{o}}^{(1)}$... $Q_{\mathfrak{o}}^{(N)}$ and the Boolean algebra they generate.   Thus, it follows from the first postulate that the set of the questions $\mathrm{III}(S)$ that can be asked to a system S has a natural structure of an orthomodular lattice containing subsets that form Boolean algebras. This is precisely the algebraic structure formed by the family of the linear subsets of a Hilbert space, which represent the yes/no measurements in ordinary quantum mechanics [Jauch 1968, Finkelstein 1969,  Piron, 1972, Beltrametti and  Cassinelli 1981].

The next problem we should face is the following. What happens if, after having asked the N question in a family $\mathfrak{q}$, the system O asks a further question Q = $Q_{N+1}$ ?    Let us introduce the second postulate.

   <u>Postulate 2 (Unlimited questions).</u>   It is always possible to acquire *new* information about a system.

If, after having asked the N questions in a family $\mathfrak{q}$, the system O asks a further question Q = $Q_{N+1}$ to the observed system S, there are two extreme possibilities: either the question Q is fully determined by the N questions in $\mathfrak{q}$, or not.  In the first case, no new information is gained.  However, the second postulate assures us that there is always a way to acquire new information.

The motivation for this postulate is fully experimental.  In fact, we know that all quantum systems (and all systems are quantum systems) have the property that even if we know their quantum state $|\psi>$ exactly, we can still "learn" something new about them by performing a measurement of a quantity O such that $|\psi>$ is not an eigenstate of O.  This is an essential *experimental* result about the world, which is coded in quantum mechanics. We expressed it in information-theoretical terms in postulate 2.  Postulate 2 is true to the extent the Planck constant is different from zero: in other words, for a macroscopic system, getting to questions that increase our knowledge of the system after having reached the maximum of our information implies measurements with extremely high sensitivity.

Since the amount of information that O can have about S is limited by postulate 1, when new information is acquired, part of the old relevant information must become irrelevant. In particular, if a new question Q (not determined by the questions in $\mathfrak{q}$), is asked, then O should loose (at least) one bit of the previous information. So that, after asking the question Q, new information is available, but the total amount of information about the system does not exceed N bits.

The next question that we should face is the extent to which the information (13) about the set of questions $\mathfrak{q}$ determines the outcome of of the additional question Q.  There are two extreme possibilities: that Q is fully determined by (13), or that it is fully independent, namely that the probability of getting a "yes" answer is 1/2. In addition, there is a range of intermediate possibilities:  The outcome of Q may be determined probabilistically by $s_{\mathfrak{A}}$. We may define, in general, as $p(Q, Q_{\mathfrak{A}}{}^{(i)})$ the probability that a "yes" answer to Q will follow the string $s_{\mathfrak{A}}{}^{(i)}$. Given two complete families of information $s_{\mathfrak{A}}$ and $s_{\mathfrak{G}}$, we can then consider the probabilities

$$p^{ij} = p(Q_{\mathfrak{G}}{}^{(i)}, Q_{\mathfrak{A}}{}^{(j)}) \qquad (17)$$

From the way it is defined, the $2^N \times 2^N$ matrix $p^{ij}$ cannot be fully arbitrary. First, we must have

$$0 \leq p^{ij} \leq 1. \qquad (18)$$

Then, if the information $s_{\mathfrak{A}}{}^{(i)}$, is available about the system, one and only one of the outcomes $s_{\mathfrak{G}}{}^{(j)}$, may result. Therefore

$$\sum_i p^{ij} = 1. \qquad (19)$$

We also assume that $p(Q_{\mathfrak{G}}{}^{(i)}, Q_{\mathfrak{A}}{}^{(j)}) = p(Q_{\mathfrak{A}}{}^{(j)}, Q_{\mathfrak{G}}{}^{(i)})$, from which we must have

$$\sum_j p^{ij} = 1. \qquad (20)$$

The conditions (18-19-20) are strong constraints on the matrix $p^{ij}$. Note that they are satisfied if

$$p^{ij} = |U^{ij}|^2 \qquad (21)$$

where U is a unitary matrix, and that $p^{ij}$ can always be written in this form for some unitary matrix U (which, however, is not fully determined by $p^{ij}$).

In order to take into account questions which are in the Boolean algebra generated by a family $s_{\mathfrak{A}}$, as for instance

$$Q_{\mathfrak{A}}{}^{(jk)} = Q_{\mathfrak{A}}{}^{(j)} \vee Q_{\mathfrak{A}}{}^{(k)}, \qquad (22)$$

we cannot consider probabilities of the form $p(Q_{\mathfrak{A}}{}^{(jk)}, Q_{\mathfrak{G}}{}^{(i)})$, because a "yes" answer to $Q_{\mathfrak{A}}{}^{(jk)}$ is less than the maximum amount of relevant information. But we may for instance consider probabilities of the form

$$p^{i(jk)i} = p(Q_{\mathfrak{G}}{}^{(i)} Q_{\mathfrak{A}}{}^{(jk)}, Q_{\mathfrak{G}}{}^{(i)}) \qquad (23)$$

defined as the probability that a "yes" answer to $Q_{\sigma}{}^{(i)}$ will follow a "yes" answer to $Q_{\sigma}{}^{(i)}$  (N bits of information) and a subsequent "yes" answer to $Q_{\mu}{}^{(jk)}$ (N-1 bits of information).   As is well known, we have (experimentally) that

$$p^{i(jk)i} \neq$$
$$p(Q_{\sigma}{}^{(i)}, Q_{\mu}{}^{(j)}) \; p(Q_{\mu}{}^{(j)}, Q_{\sigma}{}^{(i)}) + p(Q_{\sigma}{}^{(i)}, Q_{\mu}{}^{(k)}) \;\; p(Q_{\mu}{}^{(k)}, Q_{\sigma}{}^{(i)}) =$$
$$= \; (p^{ij})^2 + \; (p^{ik})^2 \qquad\qquad\qquad (24)$$

Accordingly, we can determine the missing phases of U in (21) by means of

$$p^{i(jk)i} \; = \; | \; U^{ij}U^{ji} + \;\; U^{ik}U^{ki} \; |^2 \qquad\qquad (25)$$

It would be extremely interesting to study all the constraints that the probabilistic nature of the quantities p implies, and to investigate to which extent the structure of quantum mechanics can be derived from these constraints.  One could conjecture that eqs.(21-25) could be derived solely by the properties of conditional probabilities, or find exactly the weakest formulation of the superposition principle directly in terms of probabilities: this would be an extremely strong result in the direction we are investigating. For related attempts to reconstruct the quantum mechanical formalism from the algebraic structure of the measurement outcomes, see [Mackey 1963, Maczinski 1967, Finkelstein 1969, Jauch 1968, Piron 1972].    Here, however, we do not pursue this direction.  Instead, we content ourselves with the much more modest step of introducing a third postulate.

Postulate 3 (Superposition principle).  Any question Q can be seen as part of the (Boolean) algebra generated by a complete family $\sigma$ of questions; if $\mu$ and $\sigma$ define two complete families of questions, then

$$p( \; Q_{\mu}{}^{(i)}, \; Q_{\sigma}{}^{(j)} \; ) = |U_{\mu\sigma}{}^{ij}|^2 \quad , \qquad\qquad (26)$$

where $U_{\mu\sigma}$ is a unitary matrix; for every $\mu, \sigma$ and $\rho$, we have $U_{\mu\rho}$ = $U_{\mu\sigma} U_{\sigma\rho}$; the effect of composite questions is given by eq.(25).

It follows that we may consider any question as a vector in a complex Hilbert space, fix a "basis" of questions $|Q_{\mu}{}^{(i)}\rangle$ and represent any other question $|Q_{\sigma}{}^{(j)}\rangle$ as a linear combination of these:

$$|Q_{\bar{\sigma}}^{(j)}> = \sum_i U_{\bar{\sigma}\bar{\tau}}^{ji} |Q_{\bar{\tau}}^{(i)}>. \qquad (27)$$

The matrices $U_{\bar{\sigma}\bar{\tau}}^{ij}$ could then be interpreted as generating a unitary change of basis from the $|Q_{\bar{\tau}}^{(i)}>$ to the $|Q_{\bar{\sigma}}^{(j)}>$ basis. Recall now the conventional quantum mechanical probability rule: if $|v^{(i)}>$ are a set of basis vectors and $|w^{(j)}>$ a second set of basis vectors related to the first ones by

$$|w^j> = \sum_i U^{ji} |v^i>, \qquad (28)$$

then the probability of measuring the state $|w^j>$ if the system is in the state $|v^i>$ is

$$p^{ij} = |<v^i | w^j>|^2 . \qquad (29)$$

(28) and (29) yield $p^{ij} = |U^{ij}|^2$, which is equation (26). Therefore the conventional formalism of quantum mechanics and its probability rules follow. The set $\underline{\amalg\amalg}$ (S) has the structure of a set of linear subspaces in the Hilbert space. For any yes/no question $Q_i$ let $L_i$ be the corresponding linear subset of H. The relations

$$\{ =>, \ \lor \ , \ \land \ , \text{not}, \ \perp \ \}$$

between questions $Q_i$ correspond to the relations

{ inclusion, orthogonal sum, intersection, orthogonal-complement, orthogonality }

between the corresponding linear subspaces $L_i$. The kinematics of quantum mechanics can therefore be reconstructed from the three postulates given. Note, however, that the Hilbert space on which state vectors live is not attached to a system, but to *two* systems, namely to a system-observer pair

## 3.3   Dynamics

The inclusion of dynamics in the above scheme is immediate. We simply notice that two questions can be considered as different questions if defined by the same operations but asked at different moments of time. Thus, any question Q can be labelled by the time variable t, indicating the time at which it is asked: we denote as t → Q(t) the one-parameter family of questions defined by the same procedure performed at different times. As we have seen, the set

Ш̱ (S) has the structure of a set of linear subspaces in the Hilbert space. If we postulate that time evolution is a symmetry in the theory, then the set of all the questions at time $t_2$ must be isomorphic to the set of all the questions at time $t_1$. Therefore the corresponding family of linear subspaces must have the same structure; therefore there should be a unitary transformation $U(t_2-t_1)$ such that

$$Q(t_2) = U(t_2-t_1) \ Q(t_1) \ U^{-1}(t_2-t_1). \qquad (30)$$

By conventional arguments, we then have that these unitary matrices form an abelian group and that $U(t_2-t_1) = \exp\{-i(t_2-t_1)H\}$, where H is a self-adjoint operator on the Hilbert space, the Hamiltonian. Thus, the time evolution too can be viewed as a structure on the set of the questions that can be asked to a system, or, more precisely, a structure on the family of all the questions that can be asked to the system at all times.

## 3.4   The observer observed

We now have the full formal machinery of quantum mechanics at our disposal, but with substantial interpretive differences with respect to the conventional interpretation:   There is no "state of the system" in this framework.   The issue to be addressed now is the relation between the information that different observers possess.   The important point in this regard is that the information possessed by different observers cannot be directly compared.   This is the delicate but crucial point of the entire construction.   A statement about the information possessed by O is a statement about the physical state of O; the observer O is a regular system on the same ground as any other system; thus, we may discuss his state in physical terms.   However, since *there is no absolute meaning to the state of a system*, any statement regarding the state of O is to be referred to some other system observing O.   The notion of absolute *state* of a system, and thus *a fortiori* absolute state of an observer, is not defined.   Therefore, the fact that an observer has information about a system is not an "absolute" fact: it is something that can be observed by a second observer O'.   A second observer O' can have information about the fact that O has information about S.   But recall that any acquisition of information implies a physical interaction. O' can get new information about the information that O has about S only by physically interacting with the O-S system.

We believe that a common mistake in analyzing measurement issues in quantum mechanics is to forget that precisely as an observer can acquire information about a system only by means of physically interacting with it, in the same way two observers can compare their information only by physically interacting with each other. This means that there is no way to compare "the information possessed by  O " with "the information possessed by O' ", without considering a physical interaction between the two.  Information, as any other  property of a system, is a fully relational notion.

How can a system O' have information about the fact that O has information about S?  The observer O' considers the coupled S-O system. Thus, she may ask questions to S, or to O, or to the two together. In particular she may ask questions concerning the information that O has about S.  Information is simply a property of some degrees of freedom in O  being correlated with some property of S. A question about the information possessed by O is in no way different from any other physical question.  As far as O' is concerned, knowing the physics of the S-O system means to know the set $\amalg$(S-O) of the meaningful questions that can be asked to the S-O system, and its full structure. In particular, it also means knowing how answers at time $t_1$ determine answers at time $t_2$.

A meaningful question to S is whether A is true. Denote this question as $Q_A$.  A meaningful (complete) question about O is whether his measuring apparatus is ready to (and going to) ask S the question whether A is true. Denote this question (that O' can ask to O) as $Q_{Ready}$. Another meaningful question is whether the hand of his measuring apparatus has settled to "OA". Denote this question as $Q_{OA}$. Denote the question whether the hand of his measuring apparatus has settled to "OB" as $Q_{OB}$.  Each of these questions can be asked at any time  t.

$Q_{Ready}(t)$:     O is going to measure A on S
$Q_A(t)$:     S is A
$Q_B(t)$:     S is B
$Q_{OA}(t)$:     O has the information that S is in A
$Q_{OB}(t)$:     O has the information that S is in B          (31)

We also consider a complete question $Q_{Mix}(t)$ to S such that

$$p(Q_{Mix}(t), Q_A(t)) = p(Q_{Mix}(t), Q_B(t)) = 1/2. \qquad (32)$$

and define

$$Q_{Ready,Mix}(t_1) = Q_{Ready}(t) \; \Lambda \; Q_{Mix}(t). \qquad (33)$$

Knowing the dynamics of the coupled S-O system, O' can work out the possible outcomes of questions asked at time $t_2$, given a yes answer to $Q_{Ready,Mix}(t_1)$. Assuming standard Hilbert space Hamiltonian dynamics and assuming that the coupling Hamiltonian produces a good measurement, O' will compute the probabilities

$$p( \; Q_{Ready,Mix}(t_1), \; Q_A(t_2) \; )=1/2 \qquad (34a)$$
$$p( \; Q_{Ready,Mix}(t_1), \; Q_B(t_2) \; )=1/2 \qquad (34b)$$
$$p( \; Q_{Ready,Mix}(t_1), \; Q_{OA}(t_2) \; )=1/2 \qquad (34c)$$
$$p( \; Q_{Ready,Mix}(t_1), \; Q_{OB}(t_2) \; )=1/2 \qquad (34d)$$

Thus, she has no information whether S is in A or B at time $t_2$, nor information about what the hand of the O apparatus indicates.

However, she may consider asking the question wether O knows about A or not. Let us denote this question as $Q_{"O\text{-}knows"}$. This is the conjunction of two coupled questions:

$$Q_{"O\text{-}knows"} = [Q_A \; \Lambda \; Q_{OA}] \; V \; [Q_B \; \Lambda \; Q_{OB}], \qquad (35)$$

and corresponds to the operator M described in sec.2. Note that $Q_A$, $Q_{OA}$, $Q_B$ and $Q_{OB}$ are compatible questions, so we can consider questions in the Boolean algebra they generate. The dynamics gives

$$p( \; Q_{Ready,Mix}(t_1), \; Q_{"O\text{-}knows"} \; (t_2) \; ) = 1 \qquad (36)$$

Thus, O' has information that O has information about S [equation (36)]. This of course is not in contradiction with the fact that she (O') has no information about S [equation (34a,b)], nor has she information about which specific information O has about S [equation (34c,d)]. Thus the notion "a system O has information about a system S" is a physical notion that can be studied experimentally (by a third observer), in the same way as any other physical property of a system. It corresponds to the fact that the systems S and O are correlated.

Now, the question "Do observers O and O' get the *same* answers out of a system S?" is a meaningless question. Because it is a question about the *absolute state* of O and O'. What is meaningful is to rephrase this question in terms of some observer. For instance, we could ask this question in terms of the information possessed by a further observer O", or, alternatively, by O' herself. Consider this last case. At time $t_1$, O gets information about S. O' h a s information about the initial state, and therefore has the information that the measurement has been performed. The meaning of this is that she knows that the states of the O-S system,s are correlated, or more precisely she knows that if at a later time $t_3$ she asks a question to S concerning property A, and a question to O concerning his knowledge about A (or, equivalently, concerning the position of a pointer), she will get consistent results.

From the dynamical point of view, notice that knowledge of the structure of the family of questions ш(S) that can be asked to S implies the knowledge of the dynamics of S (because ш(S) includes all Heisenberg observables at all times). In Hilbert space terms, this means knowing the Hamiltonian of the evolution of the observed system. If O' knows the dynamics of the O-S system, she knows the two Hamiltonians of O and S *and* the interaction Hamiltonian. The interaction Hamiltonian cannot be vanishing because a measurement (O measuring S) implies an interaction: this is the only way in which a correlation can be dynamically established. From the point of view of O', the measurement is therefore a fully unitary evolution, which is determined by the interaction Hamiltonian between O and S. The interaction is a measurement if it brings the states to a correlated configuration. On the other side, O gives a dynamical description of S alone. Therefore he can only use the S Hamiltonian. Since between times $t_1$ and $t_2$ the evolution of S is affected by its interaction with O, the description of the unitary evolution of S given by O breaks down. The unitary evolution does not break down for mysterious physical quantum jumps, due to unknown effects, but simply because O is not giving a full dynamical description of the interaction. O cannot have a full description of the interaction of S with himself (O), because his information is correlation, and there is no meaning in being correlated with oneself.

The reader may convince himself that even if we take into account several observers observing each other, there is no way in which contradictions may develop, provided that one does not violate the two rules:
i. there is no absolute meaning to the state of a system or the information that a system has, except within the information of a further observer.

ii there is no way a system O' may get information about a system O without physically interacting with it, and therefore breaking down (at the time of the interaction) any unitary evolution description from any observer O'' that is not including both S and O (and their interaction Hamiltonian!) in its Hilbert space description of the events.

For instance, there is no way two observers O' and O'' can get information about a system O independently from each other: one of two (say O') will have to obtain the information first. In doing so, she will interact with O at a certain time t. This interaction implies the fact that there is a non vanishing interaction Hamiltonian between O and O'. If O'' asks a question to O at a later time t', he will either have to consider the interacting correlated O-O' system, or to realize the unitary evolution of the O dynamics has broken down, due to some physical interaction he was not taking into account.

Finally, one can quantitatively study the relation between correlation of quantum states and amount of information. We will not pursue this direction here, but we present two comments in this regard. We can say in general that the property "O1" of the system O contains one bit of information about the property "1" of the system S anytime the S-O system has answered a question Q such that

$$p( Q, \quad Q_{\text{"O-knows-1"}} ) = 1 \qquad\qquad (36)$$

where

$$Q_{\text{"O-knows-1"}} = [Q_1 \wedge Q_{O1}] \vee [\text{not } Q_1 \wedge \text{ not } Q_{O1}]. \qquad (37)$$

and so on. There is an alternative (independent) characterization of amount of information, as the amount of entanglement between S and O. Using the Hilbert space formalism, we may consider the state $|\Psi\rangle$ in the tensor product Hilbert space, and the corresponding pure state density matrix $\rho = |\Psi\rangle\langle\Psi|$. If we denote the traces in the two factor Hilbert spaces by $\text{Tr}_S$ and $\text{Tr}_O$, then it is easy to see that for a non entangled state $|\Psi\rangle = |\psi\rangle_S \otimes |\phi\rangle_O$, we have

$$\text{Tr}_O \,(\text{Tr}_S \,\rho \,)^2 = 1; \qquad\qquad (38)$$

while for an entangled state of the form

$$|\Psi\rangle = \frac{1}{\sqrt{n}} \sum_{i=1,n} |\psi_i\rangle_S \otimes |\phi_i\rangle_O, \qquad\qquad (39)$$

we have

$$\text{Tr}_O \,(\text{Tr}_S \,\rho \,)^2 = 1/n. \qquad\qquad (40)$$

Thus, we may consider the amount of "entangling information" defined as

$$N = - \ln Tr_O \, (Tr_S \, \rho \,)^2. \qquad (41)$$

## 4. Critique of the concept of state

### 4.1 "Any observation requires an observer"

Let us summarize the path that we have covered.  We started from the distinction between observer and observed-system.  We have assumed (hypothesis 1) that all systems are equivalent, so that any observer can be described by the same physics as any other system. In particular, we have assumed that an observer that measures a system can be described by quantum mechanics.  We have therefore analyzed a fixed physical sequence of events **E**, from two different points of observations, the one of the observer and the one of a third system, external to the measurement.  We have concluded that two observers give different accounts of the same physical set of events.

Rather than backtracking in front of this observation, and giving up our commitment to the belief that all observers are equivalent, we have decided to take this experimental fact at its face value, and consider it as a starting point for understanding the world.  *If different observers give different descriptions of the state of the same system, this means that the notion of state is observer dependent.*  We have taken this deduction seriously, and we have considered a conceptual scheme in which the notion of absolute observer-independent state of a system is replaced by the notion of information about a system that an observer may possess.

We have considered three postulates that this information must satisfy, which summarize present experimental evidence about the world.  The first limits the amount of relevant information that a system can have; the second summarizes the novelty revealed by the experiments from which quantum mechanics has been constructed by asserting that whatever the information we have about a system we can always get new information.  The third limits the structure of the set of questions; this third postulate can probably be sharpened considerably.  Out of these postulates we have rederived the conventional Hilbert space formalism of quantum mechanics and the corresponding rules for calculating probabilities (and therefore any other equivalent formalism).

A physical system is characterized by the structure on the set $\text{Ш}(S)$ of questions that can be asked to the system. This set has the structure of the non-Boolean algebra of a family of linear subspaces of a complex k-dimensional Hilbert space. The information about S that any observer O can possess can be represented as a string s, containing an amount of information $N=\log_2 k$.

Finally, we have studied what is this "information" out of which the theory is constructed. We have shown that the fact that a system O has information about a system S means that the states of S and O are correlated, meaning that a third observer O' has information about the coupled S-O system that allows her to predict correlated outcomes between questions to S and questions to O. Thus correlation has no absolute meaning, because states have no absolute meaning, and must be interpreted as the content of the information that a third system has about the S-O couple.

Finally, since we take quantum mechanical as a complete description of the world at the present level of experimental knowledge (hypothesis 2), we are forced to accept the result that there is no "objective", or more precisely "observer independent" meaning to the ascription of a property to a system. Thus, the properties of the systems are to be described by an interrelated net of observations and information collected from these observations. Any complex situation can be described "in toto" by a further additional observer, and the interrelation is fully consistent. But there is no way to "exit" from the observer-observed global system. *Any observation requires an observer* [Maturana and Varela 1920]. In other words, we suggest that it is a matter of natural science whether or not the descriptions that different observers give of the same ensemble of events is universal or not:

Quantum mechanics is the theoretical formalization of the experimental discovery that the descriptions that different observers give of the same ensemble of events is not universal.

The concept that quantum mechanics forces us to give up is the concept of a description of a system independent from the observer providing such a description; that is, the concept of absolute state of a system. The structure of the classical scientific description of the world in terms of *systems* that are in certain *states* is perhaps incorrect, and inappropriate to describe the the world beyond the $\hbar \to 0$ limit. It is perhaps worthwhile to emphasize the fact that those considerations do not follow from the theory of quantum mechanics: they follow from a collection of experiments on the atomic world.

There are several aspects of the point of view we have discussed here that require further development. Among these: i. The reconstruction of sec.3 can certainly be made much more precise, and postulate 3 can be replaced by a weaker one. ii. The quantitative relation between amount of information and correlation between states, which we hinted at, at the end of sec.3.4, should be studied. iii. We believe that it would be extremely interesting to reconsider EPR-like issues in the light of the considerations made here; notice that there is no meaning in comparing the outcome of two spatially separated measurements, unless there is a physical interaction between the observers.

## 4.3    Relation with other interpretations

The main distinction between the present approach and Bohr's original interpretation of quantum mechanics is the fact that Bohr assumes a classical world. In Bohr's view, this classical world is physically distinct from the microsystems described by quantum mechanics, and it is precisely the classical nature of the apparatus that gives measurement interactions a special status [Bohr 1949; for a clear discussion of this point, see Landau Lifshitz 1977]. This view is pushed to an extreme by Wigner's hypothesis about the role played by consciousness in the collapse of the wave function [Wigner 1961].

Of course, from the point of view developed here, we can fix once and for all a privileged system $S_o$ as "The Observer". In this way we obtain the standard quantum mechanics interpretation. The quantum mechanical "state" of a system S is then the information that the privileged system $S_o$ has about S. From the present point of view, Bohr's choice is simply the assumption of a large (consistent) set of systems (the classical systems) as privileged observers. This is fully consistent with our view: the only difficulty is that by doing so, one becomes blind to the net of interrelations that are at the foundation of the theory, and puzzled about the fact that the physical theory treats one system, $S_o$ (classical world, consciousness, .... ), in a way which is physically different from the other systems.

Notice that the considerations in this paper do not suggest any modification to the conventional *use* of quantum mechanics: there is nothing incorrect in fixing a preferred observer once and for all. Thus, acceptance of the point of view suggested here implies continuing to use quantum mechanics precisely as is it is currently used. On the other side, this point of view (we hope) could bring some clarity about the physical significance of the strange theoretical procedure adopted in quantum mechanics: we treat a portion of the world in a different way than the rest of the world. This different treatment is, we believe, at the origin of a large part of the common unease with quantum mechanics.

The strident aspect of Bohr's quantum mechanics is cleanly characterized by Von Neumann's introduction of the "projection postulate", according to which systems have two different kinds of evolutions: the unitary and deterministic Schrödinger evolution, and the instantaneous, probabilistic measurement collapse [von Neumann 1932]. According to the point of view described here, the Schrödinger unitary evolution of the system S breaks down simply because the system interacts with something which is not taken into account by the evolution equations. Unitary evolution requires the system to be isolated, which is exactly what ceases to be true during the measurement, because of the interaction with the observer. If we include the observer into the system, then the evolution is still unitary, but we are now dealing with the description of a different observer. As suggested by Abhay Ashtekar, the point of view presented here can perhaps be characterized by a fundamental assumption prohibiting an observer to be able to give a full description of "itself" [Ashtekar 1993].

Traditional approaches that try to avoid the dual nature of Von Neumann evolution can be distinguished between a: the ones that ascribe reality to the outcome of the measurements and deny reality to the state (the wave function), and b: the ones that ascribe reality to the state and deny reality to the outcome of the measurements.

In the first class, we can include the consistent histories interpretation [Griffiths 1984, Omnes 1988] and the, related,

coherent histories interpretation [Gell-Mann and Hartle 1990]. These interpretations reduce the description of a system to the prediction of the system's responses to a set of questions that one can formulate about the system: they are quite similar in spirit to the one presented here. One should distinguish the *probability* of histories, which are observer independent, from a statement, made by an observer, that a certain history of a certain system has happened. We believe that such a statement must be meaningful within the theory if we are to make contact with real observations. In appendix A, we argue that within the history interpretations different observers *do* make different statements about the same sequence of events. One may then want to treat the observer that makes a specific statement about *one* outcome as a quantum system, and understand the *physical* relation between different descriptions given by different systems. This is precisely what we tried to achieve here. Thus, the histories interpretations are consistent with the analysis developed here. What we add here is increased attention to the process through which the observer-independent probabilities, attached to families of histories, may be related to *actual* descriptions.

Similar considerations hold for any other interpretation that views the outcome of the measurement as the only "real" aspect of the theory [van Fraassen 1991], or as an actualization of a potentialities [Shimony 1969, Fleming 1992]. Indeed, quantum mechanics is characterized precisely by the fact that the notion of outcome of a measurement, or the notion of properties being ascribed to a system only at a certain time t, only make sense at the price of singling out *preferred* moments of times, in which the potentiality becomes actual. What is it that determines this moment of time? As A Shimony put it: "It remains an open problem, however, ... to determine under what circumstances the actualization of potentialities generically take place." [Shimony 1969, quoted in Fleming 1992]. The observation at the basis of this paper is that this moment of time in which the potentialities become actual, or the wave function collapse, is simply determined by the moment in which a specific physical system not including the observed system interacts with the observed system. But then the moment in which potentialities become actual, or in which the empiricist's "data" come into existence, is completely observer dependent in quantum mechanics.

Examples of interpretations in the group b are the ones in which the measurement process is replaced by some hypothetical physical process that violates the linear Schrödinger equation [Ghirardi Rimini and Weber 1986, Penrose 1989]. Those interpretation are radically alternative to the present approach, since they violate hypothesis 2. Other examples in the group b are the many world interpretation [Everett 1957, Wheeler 1957, DeWitt 1970], and its variants. If the "branching" of the wave function in the many world interpretation is considered as a physical process, it raises, we believe, the very same sort of difficulties as the von Newmann "collapse" does. When does it happen? Which systems are measuring systems that make the world branch? These difficulties of the many-world interpretation have been discussed in the literature [See Earman 1986]. If, on the other side, we forget branching, and we keep evolving the wave function under the unitary evolution, we obtain the description of the world provided by a fully external observer, and the problem is to interpret the observation of the "smaller" observers. This is precisely the problem addressed in this paper. Thus, also the many worlds interpretation can be taken as a starting point of the present analysis - although an opposite starting point than the histories interpretations.

Thus, from the point of view discussed here, Bohr's interpretation, consistent and coherent histories interpretations, as well as the many worlds interpretation, are all quite literarily correct, albeit incomplete.

Outside physics, the idea of the importance of the observer in any description of any system [Maturana and Varela 1980, von Foerster 1982] is today almost a commonplace in disparate areas of thinking [as an extreme example: Bragagnolo Cesari and Facci 1993]. In philosophy, both the post-neopositivist tradition and (much more radically) continental philosophy (as an example: [Gadamer 1989]), have more and more emphasized the need of referring any description of an object to the describing observer. It is curious that quantum mechanics, which contributed so much to inspire these views, has not yet taken full advantage of them.

The point of view closest to the one presented here is perhaps Heisenberg's. Heisenberg's insistence on the fact that the lesson to be taken from the atomic experiments is that we should stop thinking of the "state of the system", has been obscured by the subsequent limpid definition of the theory in terms of states given by Dirac.  Here, we have taken Heisenberg's lesson to its extreme consequences.

Our greatest debt, however, is to John Wheeler.  John's insistence through the years that we have not yet understood "the Quantum", and his fascinating indications on the connection between information theory and quantum mechanics ("It from Bit") have directly inspired the present work.

**Acknowledgements**
The ideas at the root of this work emerged from: i. A seminar on the interpretation of quantum mechanics run by Bob Griffiths at Carnegie Mellon University (1993); ii. A seminar on the same subject run by John Earman at Pittsburgh University (1992); iii. Louis Crane's ideas on quantum cosmology; iv. Discussions with Abhay Ashtekar, Simonetta Frittelli, Al Janis, Ted Newman, Lee Smolin;  v. A comment made by Bob Geroch; vi. The teachings of Paola Cesari on the importance of taking the observer into account.  It is a pleasure to thank them all. In addition, I'd like to thank Al Janis for a carefull reading of this manuscript. and for many useful criticisms and comments.

**Appendix A**

This appendix can be considered as the continuation of section 2.3. We consider here other possible objections to the observation that different observers give different descriptions of the same sequence of events.

*Objection 1. The account*  (4) *is the correct one.  There is no collapse, nor classical properties of system, but only quantum states. The description (1) is not correct, because "the wave function never really collapses".*
If so, then the observer O' cannot measure the A property either, since (because of the fundamental assumption) "it cannot make the function collapse either"; thus the system S doesn't *ever*  have the property A.  But we describe the world in terms of "properties" that the systems have, namely values that observables take, not in terms of states in the Hilbert space. In a description of the world in which the wave function never collapses, the systems never have definite properties and we do not see how to match the description with any observation. [Albert 1992] If we say "When 'we' observe the system, then we 'see' definite properties", then we ascribe to ourselves a non physical behavior, which is what we are refusing to do. Alternatively, we accept that 'we' can be described (in principle) quantum mechanically, and thus we are forced back to the fact that different observers give different descriptions of the same events.

*Objection 2. The account* (1) *is the correct one. Outcomes of the measurements are the only true content of the theory. At time $t_2$, O does measure A, then the value of A is absolutely determined, albeit unknown to O', (who could measure it later).*

This objection is, in a sense, opposite to objection 3. As formulated, the objection is just wrong: We cannot assume that at time $t_2$ the value of A is absolutely determined, albeit unknown to O' because at time $t_3$ O' may measure an observable that does not commute with A, and the outcome of which is affected by interference terms between the two terms in the right hand side of (4). According to the objection, these interference terms should not exist, while according to quantum mechanics they are observable. This objection confuses quantum mechanical superposition with statistical states: the r.h.s of (4) is a pure state.

*Objection 2 bis. The account* (1) *is the correct one. The interference terms mentioned above become extremely small if O is macroscopic (because of the decoherence effect). If they are small enough, they are unobservable, and thus A becomes an absolute property of S, which is true and absolutely determined, albeit unknown to O', who could measure it anytime, and would not see interference effects.*

Again, this is definitely wrong, because according to this hypothesis a property A of S can become an absolute property at time $t_2$, or not, according to which subsequent properties of S the observer (O') considers. (We do not prove this here.) This precisely the reason for which coherent histories interpretations can (consistently) assign probabilities to histories, and not to single outcomes of measurements within a history.

*Objection 3. What is absolute and observer independent is the probability of a sequence $A_1, ... A_n$ of property ascriptions (such that the interference terms mentioned above are extremely small -*

*decoherence), this probability is independent from the existence of any observer measuring these properties.*

This is certainly correct, and, in fact, this observation is at the root of the coherent histories interpretations of quantum mechanics [Griffiths 1984, Omnes 1988, Gell-Mann and Hartle 1990]. But this interpretation confirms our conclusion above that different observers give different accounts of the same sequence of events, for the following reason. The beauty of the histories interpretations is the fact that the *probability* of a sequence of outcomes does not depend on the observer, precisely as it doesn't in classical mechanics. One can be content with this aspect of the theory and consistently stop here. However, one can also investigate how a concrete observer (a physicist, a machine) checks that the predicted probabilities are correct, or uses the theory. In using the histories formalism we do not have to make a wave function collapse and we can forget the observer; but if we do want to compare the theory's predictions with a set of data, or a *physical account of a set of real physical events that have happened*, then we must recall that this physical account depends on the *choice* of a consistent family of histories used to describe this sequence of events. The observer makes a choice in picking up the family of alternative histories in terms of which to describe the system. Unlike classical mechanics, this choice is such that the same property can be true or not, according to the family chosen [Griffiths 1993]. Now, since the observer too is a physical system, this choice must be dictated by the kind of physical interaction that occurs between the observer O and the system S. Unless we assume that the observer is an unphysical entity, we then are free to consider how a second observer would describe the coupled S-O system, and we are back to the problem above: the descriptions of the same sequence of events given by the two observers can be different. Classes of observers may agree on sets of outcomes (or interpret the differences as statistical ignorance), but there may always be other observers, perhaps observing the sequence of events *and* O, who have chosen a different alternative family of consistent histories to describe the same sequence of events. Note that what is different is not the probabilities, but the actual description of what happened in a concrete instance. *Thus, the consistent histories interpretation confirms the conclusion that if an observer O gives a description of a sequence of events, another observer may give a different description of the same sequence.*